\RequirePackage{fix-cm}
\documentclass[smallextended]{svjour3}          
\smartqed  
\usepackage{graphicx}
\usepackage{a4wide}
\usepackage{color}
\usepackage{epsfig}
\usepackage{amsmath}
\usepackage{amsfonts}
\usepackage{amssymb}
\usepackage{latexsym}
\usepackage{theorem}
\usepackage{setspace}

\begin{document}

\title{A Spatial Markov Chain Cellular Automata Model for the Spread of Viruses} 

\author{F.J. Vermolen
}

\authorrunning{Short form of author list} 

\institute{F.J. Vermolen\at
              Delft Institute of Applied Mathematics \\
	      Delft University of Technology \\
	      Delft, The Netherlands \\
              \email{F.J.Vermolen@tudelft.nl} 
}

\date{Received: date / Accepted: date}

\maketitle

%
%
\sloppy
\abstract{We consider a Spatial Markov Chain model for the spread of viruses. The model is based on the principle to represent a graph connecting
nodes, which represent humans. The vertices between the nodes represent relations between humans. In this way, a graph is connected in which the
likelihood of infectious spread from person to person is determined by the intensity of interpersonal contact. Infectious transfer is determined by chance.
The model is extended to incorporate various lockdown scenarios.}

\section{Introduction}
\label{sec:1}
At the time of writing, April 2020, the global human population is being hit by the Corona virus, in which some of the 
infected individuals can become seriously ill so that nursing at the intensive care unit is necessary, or that death occurs.
The disease, COVID19, is often characterised by flu-like symptoms, which in some cases lead to excessive fever or even to lung
inflammations. One of the serious problems regarding this disease is the high infection rate from person to person. 
In order to protect as many people as possible against the virus, many governments, including in Belgium, Britain and 
France, implement lockdown strategies. Other countries in Europe choose for herd immunity. Examples of such
countries are Netherlands and Sweden. Both the Netherlands and Sweden show, despite their small population
sizes, a relatively large mortality \cite{politico}. Despite their adopting of a firm lockdown strategy, Italy and Spain regret
a high number of mortalities as a result of the Corona virus. Some criticasters claim that these high mortality rates
are a consequence of having implemented the lockdown strategy only at a late stadium. Germany, on the other
hand, has installed a lockdown strategy, and has registered many cases of infected people, however, fortunately
Germany is only effected by a relatively
low mortality. A reason for this low mortality rate could be the intensive testing policy that has been carried out by the 
Germans. Sweden and Netherland could possibly blame their high mortality numbers to the fact that testing is done
on a small scale only. Only caregivers and very seriously ill people are tested in Sweden and Netherlands.

In order to predict the dynamics of the spread, death rate and recovery rate of the Corona virus, many different 
strategies are used. A very common model is the so-called SIR model, see \cite{sir} for the original paper.
More modern elaborations on the SIR principle have been presented in \cite{Getz} and \cite{Allen}. In particular, 
the model in \cite{Allen} bears some similarities with the approach that is presented in the current paper.
This model simulates a homogeneous population
that is exposed to a virus. It contains a susceptible, resistent (recovered) and infected fraction of the population.
Many more advanced models are variations on this strategy. One attempts to include spatial spread by the incorporation
of diffusion terms, which are justified by random (unpredictable) migration and interaction of individuals. Other extensions
are based on the incorporation of networks, which allows so-called jump processes so that airborne communication
can be taken into account. The models described in \cite{Trapman} distinguish several challenges for network modelling
in epidemics. The current model elaborates on the influence of the topology of the network on the evolution of the 
epidemic.

The current model has a stochastic nature. We consider a list of individuals, which can be related to each other. Each 
connection between two individuals contains a likelihood that a two individuals infect each other. Of course contact
between individuals does not always lead to infection, and hence here a stochastic process is considered. The likelihood
that each individual infects another is determined by the intensity of the contacts that the individuals have. Next to being
infected, recovery is incorporated and once an individual has recovered, then it is assumed that the individual is immune
to the disease. Since COVID19 can be a lethal disease in some cases, death has been incorporated as well. The model
has been extended for modelling lockdown policies that certain governments have adopted. One of the advantages of
the current approach is the small number of input parameters needed. A further innovation is the uncertainty quantification
and the statistical assessment of the results.

Section 2 explains the mathematical model, which builds further on \cite{Polonen}. Section 3 contains the numerical
implementation and the implementation of lockdown scenarios. Section 4 contains computer simulations and Section 5 
ends up with conclusions.
\section{The Mathematical Model}
\label{sec:2}
We consider a graph with nodes and vertices. The nodes represent individuals that can be in four states: susceptible, 
infected, resistent (or recovered), dead. If a person is susceptible, then this individual can be infected. Once the individual
is infected, then, the person can either recover or die. If (s)he recovers then this person is assumed to be resistent. If a person
is susceptible, dead or resistent, then (s)he will not spread the virus to other people (although this assumption may be subject
to discussion because a non-infected could spread the virus by the hands or other objects, but this effect is neglected in the
current modelling). The interpersonal relations are represented by vertices in the graph. Each connection between two
nodes represent a connection. The connection is subject to an intensity, which represents the frequency that two individuals
physically interact. This intensity determines the likelihood that, if one of the two individuals is infected, the disease is
transferred from one to the other individual. Furthermore, infected individuals may recover or die.

Mathematically, we consider the following: We have $n$ individuals and a vector of length $n$, where each entry in this
vector contains the state of individual. This vector is denoted by ${\bf v}$, where the value of $v_i$ contains the integer 
states: $v_i \in \{1,2,3,4\}$, where $v_i = 1$, $v_i = 2$, $v_i = 3$ and $v_i = 4$, respectively, correspond to the
susceptible, infected, resistent and death states. All individuals are connected to other individuals by vertices between
nodes (or individuals). The connection between person $i$ and $j$ is denoted by $a_{ij}$, where $a_{ij} = 0$ represents
the case that individuals $i$ and $j$ have no physical contact. The entries $a_{ij}$ are assembled into the adjacency 
matrix $A$. Large values of $a_{ij}$ represent the intensity of the contacts. We will consider the dynamics of the spread
of the disease in the coming subsections.
\subsection{The transfer of the virus from individual to individual}
First we consider the transfer from individual $i$ to individual $j$. Suppose that $a_{ij} \in [0,1]$, then these two individuals are
in physical contact. Suppose that person $i$ is infected and that person $j$ is susceptible. Then we assume that the 
infection of $j$ is a memoryless stochastic event and that it follows an exponential distribution with infection probability
rate $\lambda_{ij}$, given a time interval $\tau$, then the likelihood that person $i$ infects person $j$ is given by
\begin{equation}
P(v_j(t + \tau) = 2 | v_j(t) = 1) = \int_t^{t + \tau} \lambda_{ij}(s) \exp{(-\lambda_{ij}(s)(s-t))} ds.
\end{equation}
Furthermore, we have 
\begin{equation}
P(v_j(t + \tau) = 1 | v_j(t) = 1) = 1 - \int_t^{t + \tau} \lambda_{ij}(s) \exp{(-\lambda_{ij}(s)(s-t))} ds,
\end{equation}
and hence 
\begin{equation}
P(v_j(t + \tau) \in \{3,4\} | v_j(t) = 1) = 0.
\end{equation}
Death of individuals can also follow from other causes than Corona, but these causes are not incorporated in the current
modelling. The transfer probability rate to node $j$ is determined by the infection state of the neighbour nodes of $j$, let the
set of neighbours of node $j$ be given by
\begin{equation}
\mathcal{N}_j ;= \{k \in \{1,\ldots,n\} ~:~ a_{kj} > 0\}.
\end{equation}
For the transfer probability rate $\lambda_{ij}$, we have
\begin{equation}
\lambda_{ij}(t) = a_{ij}(t) \hat{\lambda},
\end{equation}
where we incorporated the time dependence of the intensity of contacts.

Next we consider all neighbours of node $i$ in the susceptible state ($v_i = 1$), that is the set $\mathcal{N}_i$, consider the set of infected neighbours of
node $i$, that is
\begin{equation}
\mathcal{N}^I_i (t) := \{ k \in \mathcal{N}_i ~ : ~ v_k(t) = 2\}. 
\end{equation}
Then the likelihood that node $i$ will not transfer into the infected state is given by
\begin{equation}
P(v_i(t+\tau) = 1 | v_i(t) = 1) = \prod_{j \in \mathcal{N}^I_i(t)} (1 - \int_t^{t + \tau} \lambda_{ij}(s) \exp{(-\lambda_{ij}(s)(s-t))} ds).
\end{equation}
This implies that 
\begin{equation}
P(v_i(t+\tau) = 2 | v_i(t) = 1) = 1 - \prod_{j \in \mathcal{N}^I_i(t)} (1 - \int_t^{t + \tau} \lambda_{ij}(s) \exp{(-\lambda_{ij}(s)(s-t))} ds). 
\end{equation}
If we simplify the above expression such that $\lambda_{ij}(s) = \lambda_{ij}(t)$, then the above expression becomes
\begin{equation}
P(v_i(t+\tau) = 2 | v_i(t) = 1) = 1 - \prod_{j \in \mathcal{N}^I_i(t)} \exp{(-\lambda_{ij}(t)\tau)} = 1 - \exp{(- \tau \sum_{j \in \mathcal{N}^I_i} \lambda_{ij}(t))}. 
\end{equation}
Combined with the adjacency matrix, we obtain
\begin{equation}
P(v_i(t+\tau) = 2 | v_i(t) = 1) = 1 - \exp{(- \tau \sum_{j \in \mathcal{N}^I_i(t)} a_{ij}(t) \hat{\lambda})}. 
\end{equation}
This implies that the effective transfer probability rate for node $i$ is given by
\begin{equation}
\lambda^{{\text eff}}_i = \hat{\lambda} \sum_{j \in \mathcal{N}^I_i(t)}a_{ij}(t).
\end{equation}
This is summarised in Theorem 1, of which a similar version was proved in \cite{Polonen} \\[2ex]
{\bf Theorem 1:} {\em Let node $i$ possess neigbours $\mathcal{N}^I_i(t)$ that are infected. Then, assuming the adjacency matrix not to change
during the time interval $(t,t+\tau)$, the effective probability rate in the
exponential distribution for node $i$ to become infected is given by }
$$\lambda^{{\text eff}}_i = \hat{\lambda} \sum_{j \in \mathcal{N}^I_i(t)} a_{ij}(t).$$
Next, we consider the transition from being infected (possibly ill) to the recovered (resistent) or dead state.
\subsection{Transition to recovery and death}
People that are in the infected state await two different scenarios: recovery with being resistent or death. Some people recover very quickly after having had (very) mild or 
even no symptoms, whereas other people need a long time to recover or, worse, even pass away. In the current modelling, it is assumed that the recovery time follows
an exponential distribution, that is
\begin{equation}
P(v_i(t+\tau) \in \{3,4\} | v_i(t) = 2) = \int_t^{t + \tau} \mu \exp{(-\mu (s - t))} ds,
\end{equation}
where $\mu > 0$ represents the probability rate for recovery. It has been assumed that $\mu$ is constant. It is realised that $\mu$ may be subject to temporal
changes due to improvements of medical therapies against the disease. The mean recovery time from the moment that the patient was infected is determined by
\begin{equation}
T_r = \frac{1}{\mu}.
\end{equation}
Let us assume that the mortality likelihood after infection is given by $\alpha$, then, we assume that all people die who did not heal after a length of time
interval $T_d$ after infection. Hence suppose that individual $i$ is infected at time $t$, then $T_d$ is determined by
\begin{equation}
1 - \int_{t}^{t + T_d} \mu \exp{(-\mu (s - t))} ds = \alpha \Longleftrightarrow \mu \int_0^{T_d}  \exp{(-\mu s)} ds = 1 - \alpha.
\end{equation}
This implies that 
\begin{equation}
1 - \exp{(-\mu T_d)} = 1 - \alpha \Longleftrightarrow T_d = -\frac{1}{\mu} \log(\alpha) = -\log(\alpha) T_r.
\end{equation}
Here we use the natural logarithm.
Note that $\alpha$ represents a probability, which does not exceed one, and in particular it is a small number in the order of at most 2--3 \% = 0.02 -- 0.03 at most.
For $\alpha = 0.05$, we have $T_d \approx 3 T_r = \frac{3}{\mu}$.
We assume that patients that have been ill during more than a time-interval $T_d$ have had so much damage to their vital organs (lungs and kidney) that they 
die. Mathematically, we can write if patient $i$ got infected on time $t_0$, that is $\displaystyle{t_0 = \min_{t \ge 0} \{v_i(t) = 2}\}$,
\begin{equation}
v_i(t_0+\theta) = 
\begin{cases}
3, \qquad  \text{ if } \theta < T_d, \\ 
4, \qquad  \text{ if } \theta \ge T_d.
\end{cases}
\end{equation}
\section{Computational Implementation}
The implementation has been done in Python. The current preliminary computations involve a simplified square topology, in which each each node has at most four 
connections. It is easy to revise this. We use a uniform transmissibility probability rate $\hat{\lambda}$ to obtain the probability that the node changes from susceptible to 
infected during the time-step $\tau$. To model transmission, the effective transmission probability rate is computed by the
use of the adjacency matrix. Subsequently for each susceptible node a random number, $\xi$, from the standard uniform distribution (between zero and one) is sampled,
that is $\xi \sim U(0,1)$. If 
the number is smaller than the probability of transmission from susceptible to infected then the state is changed from susceptible to infected, that is
$v_i$ is changed from 1 to 2, that is
$$
v_i(t + \tau) = 
\begin{cases}
2, \qquad \xi < P(v_i(t+\tau) = 2|v_i(t) = 1), \\ \\
1, \qquad \xi > P(v_i(t+\tau) = 1|v_i(t) = 1).
\end{cases}
$$
Otherwise, it stays in the susceptible state. 

The same is done for the transition from the infected state to the resistent or dead state. However, we keep track of the time-interval that a node has remained by adding
adding the time-step $\tau$ to the time-interval that a node is in the infected state. If the total length of the time interval that a nodal point stays in the infected state 
exceeds the length $T_d$, then the node is moved to the dead state. As long as this has time-interval has not been exceeded, the node is either kept as it is or 
transferred to the resistent state analogously to the treatment of susceptible nodes.

Since interpersonal contacts are often fluctuating (like going to shops, meeting friends, working, etc), we use randomised values for the adjacency matrix
$a_{ij}$, that is, considering person $i$:
$$\text{For } s \in (t,t+\tau)~:~a_{ij}(s) \sim U(0,1), \qquad \text{ if } j \in \mathcal{N}_i.$$

In the case of lockdown, the adjacency matrix is premultiplied by a factor, $\beta$, whose value ranges between zero and one. Small values of the pre-factor $\beta$
represent severe lockdown policies. That is the adjacency matrix becomes re-defined by
$$
\hat{A}  = \beta(t) A,
$$
and in all expressions given earlier, $A$ and its entries are replaced with $\hat{A}$ and its corresponding entries $\hat{a}_{ij} = \beta(t) a_{ij}$. Note that 
the lockdown policy depends on $t$, and therefore $\beta = \beta(t)$, where $\beta:\mathbb{R}^+ \longrightarrow [0,1]$.

In order to compute the fractions of susceptible, dead, infected and resistent people, we introduce the standard Kronecker Delta Function:
$$
\delta_{p,k}: \mathbb{N} \times \mathbb{N} \longrightarrow \{0,1\}~: \qquad \delta_{p,k} = 
\begin{cases}
1, \qquad \text{ if } k = p, \\ 
0, \qquad \text{ else.}
\end{cases}
$$
The fraction of individuals in state $p \in \{1,2,3,4\}$ is given by
$$
f_p(t) = \frac{1}{n} \sum_{j = 1}^n \delta_{p,v_j(t)}.
$$
This definition reproduces that $\displaystyle{\sum_{p \in \{1,2,3,4\}} f_p = 1}$. The computations are terminated as soon as the number of
infected people equals zero.

In order to eradicate the virus, it is needed that the number of newly infected people is smaller than the number of people that recover from the virus.
The likelihood that susceptible individual $i$ gets infected during the time interval $(t,t+\tau)$ is given by 
$$
P(v_i(t+\tau) = 2 | v_i(t) = 1) = \int_{t}^{t+\tau} \lambda_{\text{eff}}^i \exp(-\lambda_{\text{eff}}^i (s-t)) ds,
$$
which implies that the total number of people getting infected during the interval $(t,t+\tau)$ can be estimated by
$$
N_{\text{inf}}(t) = \sum_{j \in \{1,\ldots,n\}} \delta_{1,v_j(t)} P(v_i(t+\tau) = 2 | v_i(t) = 1) = 
\sum_{j \in \{1,\ldots,n\}} \delta_{1,v_j(t)} \int_{t}^{t+\tau} \lambda_{\text{eff}}^i \exp(-\lambda_{\text{eff}}^i (s-t)) ds.
$$
The number of people that either die or recover from the virus during the interval $(t,t+\tau)$ is given by
$$
N_{\text{noninf}}(t) = \sum_{j \in \{1,\ldots,n\}} \delta_{2,v_j(t)} \mu = n f_2(t) \mu.
$$
Since the virus decays if $N_{\text{inf}}(t) < N_{\text{noninf}}(t)$, which implies
$$
\sum_{j \in \{1,\ldots,n\}} \delta_{1,v_j(t)} \int_{t}^{t+\tau} \lambda_{\text{eff}}^i \exp(-\lambda_{\text{eff}}^i (s-t)) ds < n f_2(t) (1 - \exp(-\mu \tau)).
$$
An upper bound for the number of individuals that get infected is determined by
$$
N_{\text{inf}}(t) < n f_1(t) (1 - \exp(-\nu \cdot \max_{(i,j) \in \{1,\ldots,n\}^2} |a_{ij}| \hat{\lambda} \tau)),
$$
where $\displaystyle{\nu = \max_{j=1\ldots} |\mathcal{N}_j|}$, which is the maximum number of neighbours of any individual. Hence a sufficient
condition for a decay of the virus is obtained for
$$
f_1(t) (1 - \exp(-\nu \cdot \max_{(i,j) \in \{1,\ldots,n\}^2} |a_{ij}| \hat{\lambda} \tau)) < f_2(t) (1 - \exp(-\mu \tau)).
$$
Taylor's Theorem applied to both sides in the above equality implies
\begin{equation}
\nu\cdot \max_{(i,j) \in \{1,\ldots,n\}^2} |a_{ij}|  \lessapprox \frac{f_2(t) \mu}{f_1(t) \hat{\lambda}},
\label{compliance}
\end{equation}
which is a sufficient condition to have the epidemic decay. This implies that the number of contacts between individuals should be decreased (decrease $\nu$) or the 
intensity of contacts should be decreased (decrease of $\max_{(i,j) \in \{1,\ldots,n\}^2} |a_{ij}|$) in order to have the epidemic vanish.
\section{Computer Simulations}
In the basic configuration, we consider a rectangular arrangement of $100 \times 100$ nodes. Each node has five neighbours, except for the nodes on the boundary,
which have three neighbours. Initially all nodes are in susceptible state, except for one nodal point in the
centre of the domain. The numerical input values are given in Table 1. 
\begin{center}
\begin{tabular}{ c c c }
 Parameter & Value & Unit \\ 
 \hline 
 \hline
 $n_x \times n_y$ & $100 \times 100$ & - \\  
 $\Delta t$ & $0.1$ & day \\
 $\hat{\lambda}$ & $5 \cdot 10^{-3}$ & day$^{-1}$ \\    
 $\mu$ & $2.5 \cdot 10^{-1}$ & day$^{-1}$ \\
 $T_d$ & $5$ & day \\
\end{tabular}
\end{center}
\begin{center}
Table 1: Input values for the basic run
\end{center}

We stress that we have used hypothetic values in the current simulations and hence the simulations do not reflect reality.
Calibration of the model is done in a future step. The current paper describes a feasibility study with preliminary simulation outcomes.
\subsection{A Basic Run}
\begin{figure}[b]
\sidecaption
\includegraphics[scale=.5]{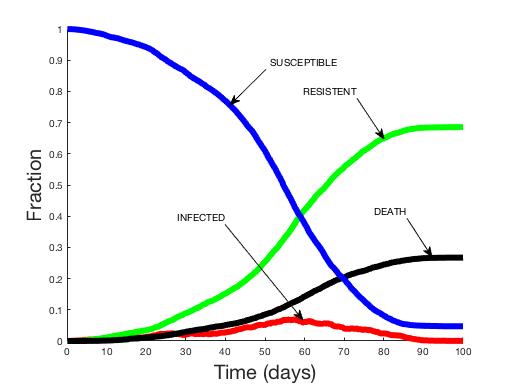}
%
%
\caption{The fractions of susceptible, infected, resistent and dead people as a function of time}
\label{fig:1}       
\end{figure}

First we consider a run in which no lockdown policy is implemented. Figure 1 depicts the fractions of susceptible,
infected, resistent and dead people. The results have been plotted in Figure \ref{fig:1}. 
The first three parameter more or less follow the dynamics of the standard SIR models for the dynamics of an epidemic. 
The number of susceptible people decreases monotonically down to an end value, whereas the infected people increase during the early stages and decrease down to
zero at the latest stages. This number has to decrease down to zero eventually since all infected people either recover and hence become resistent, or die. The curve
for the infected people shows a slight non-monotonic behaviour due to the stochastic processes that sometimes allow more people to get infected
than people transfer to death or recovery or vice versa. Furthermore
the number of resistent also grows monotonically since resistent people cannot be infected again, and in the model they will not die of corona. In the first stages, the
number of resistent and dead people follows more or less a logistic curve and so does the susceptible fraction. It can be seen that the epidemic has disappeared 
after a bit more than 100 days.
\begin{figure}[b]
\sidecaption
\includegraphics[scale=.3]{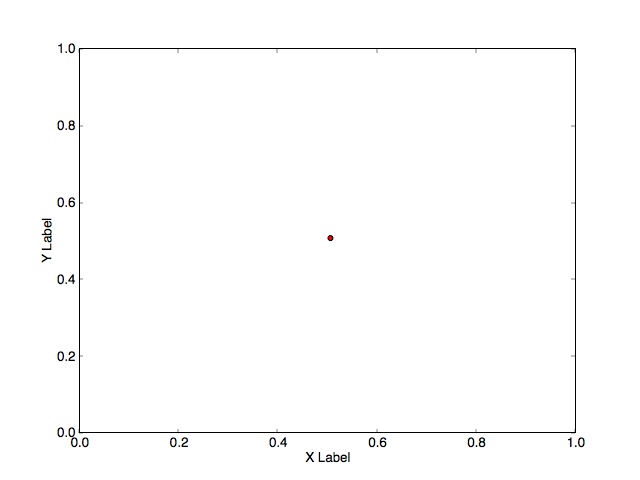}
\includegraphics[scale=.3]{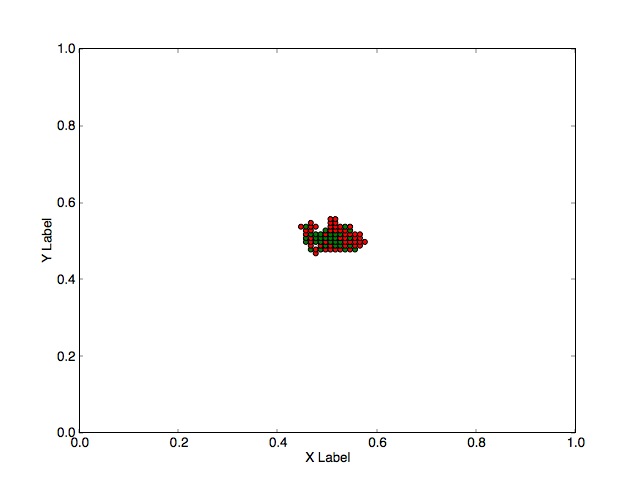}
\includegraphics[scale=.3]{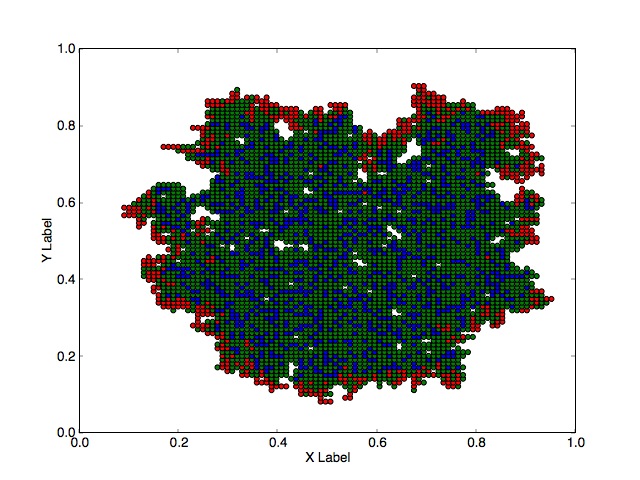}
\includegraphics[scale=.3]{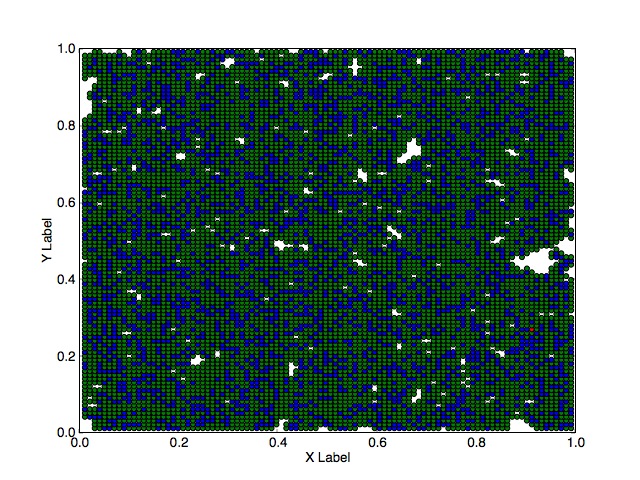}
%
%
\caption{Spatial plots of progression of the virus through the community. White, red, green and blue regions, respectively, represent
susceptible, infected, resistent and dead individuals. Snapshots are after 1 day, 10 days, 50 days and 110 days}
\label{fig:2}       
\end{figure}
In Figure \ref{fig:2}, we plot the spatial distribution of the same run as in Figure \ref{fig:1}. The red dots represent the infected individuals, the green dots are the
resistent people, and the blue dots correspond to the dead people. The white dots are susceptible that have not yet been in contact with the virus. It can be seen 
that there is an infection wave that spreads through the community over time. This wave is followed by dead and recovered people. Once a person is surrounded by
recovered people and no longer interacts with infected people, then this person will never become infected. In this sense, this models protection of some
people by resistent people. A weakness of the model is that in the current calculations, someone can be surrounded by dead people only. It is thought that these
effects are 'second order effects' and therefore, we will disregard this issue for the time being. In future studies, this issue will be quantified.
\begin{figure}[b]
\sidecaption
\includegraphics[scale=.5]{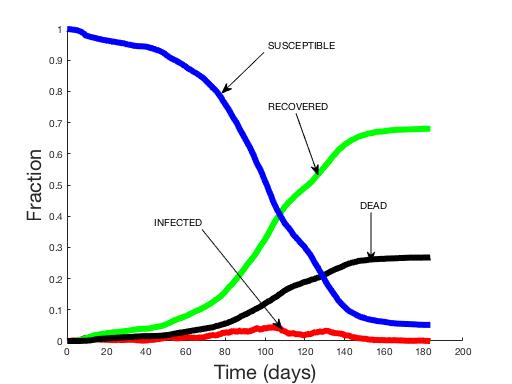}
%
%
\caption{The fractions of susceptible, infected, resistent and dead people as a function of time. A weak lockdown has been applied}
\label{fig:3}       
\end{figure}
\subsection{Weak and strong lockdown scenarios}
We consider a run in which we have a weak lockdown, that is we use
$$
\beta(t) = 
\begin{cases}
\tilde{\beta} = 0.5, \qquad \text{ for } t \in T_{ld} = (10,50), \\ 
1, \qquad \text{ else }.
\end{cases}
$$
\begin{figure}[b]
\sidecaption
\includegraphics[scale=.35]{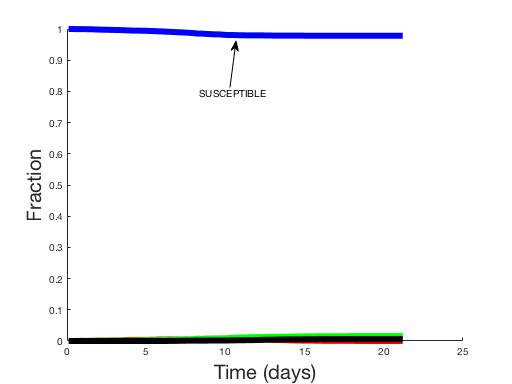}
\includegraphics[scale=.35]{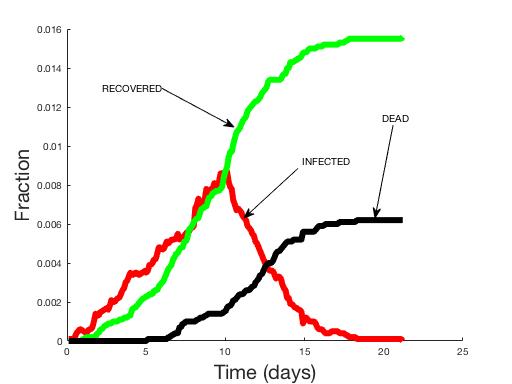}
%
%
\caption{The fractions of susceptible (left), infected, resistent and dead people (right) as a function of time. A firm lockdown has been applied.
Note the change of scale}
\label{fig:4}       
\end{figure}
This case corresponds to halving the amount of physical social interaction.
It can be seen that such a lockdown does not change the number of casualties (dead) or resistent people in the end with respect to the case in which one does not
have any policy. The only advantage is that it the number of infected people is spread over a larger time-span so that the peak of patients at the hospitals is 
smoother. The results have been plotted in Figure \ref{fig:3}. A more severe lockdown scenario, in which the amount of social interaction is reduced to
20 \% as follows
$$
\beta(t) = 
\begin{cases}
\tilde{\beta} = 0.2, \qquad \text{ for } t \in T_{ld} = (10,50), \\ 
1, \qquad \text{ else }.
\end{cases}
$$
During the time interval $t \in (10,50)$, the maximum value of the adjacency matrix is equal to $\beta(t)$, and hence a smaller minimum makes it easier
to satisfy condition (\ref{compliance}).
\begin{figure}[b]
\sidecaption
\includegraphics[scale=.33]{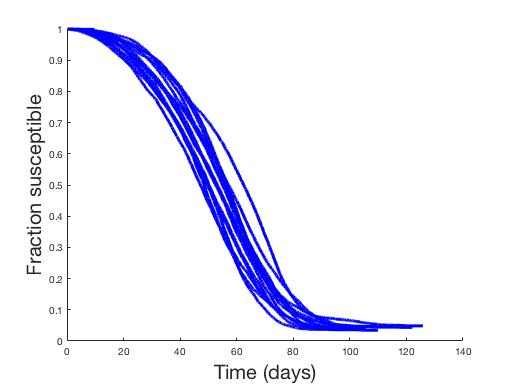}
\includegraphics[scale=.33]{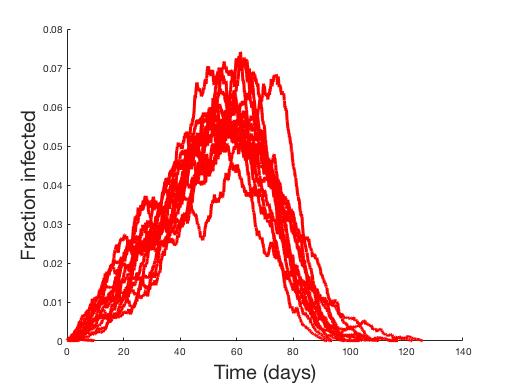} \\
\includegraphics[scale=.33]{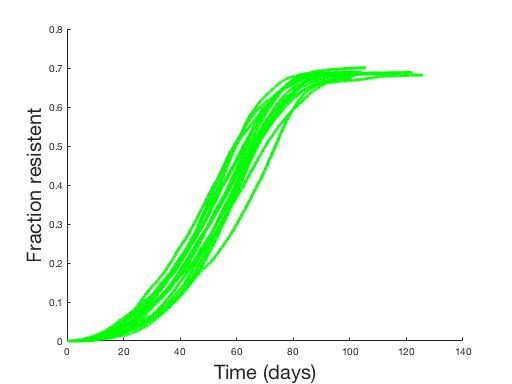}
\includegraphics[scale=.33]{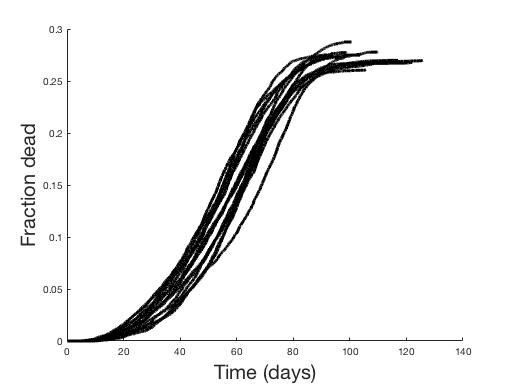}
%
%
\caption{The fractions of susceptible, infected, resistent, and dead people as a function of time without any lockdown policies. 
Twenty runs are shown. The same input data as in Figures 1 and 2 from Table 1 have been used}
\label{fig:5}       
\end{figure}
\begin{figure}[b]
\sidecaption
\includegraphics[scale=.33]{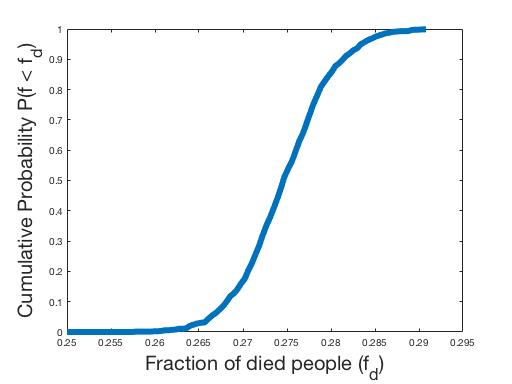}
\includegraphics[scale=.33]{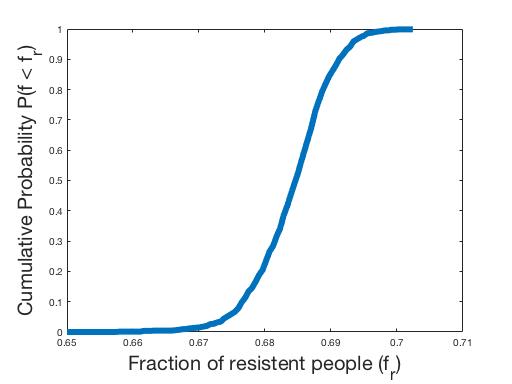} \\
\includegraphics[scale=.33]{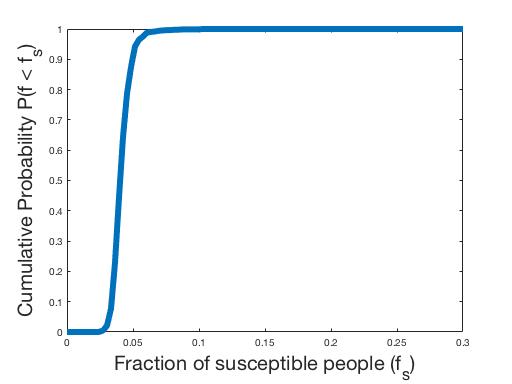}
\includegraphics[scale=.33]{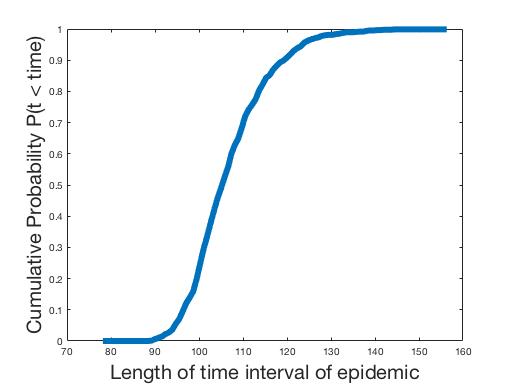}
%
%
\caption{Cumulative probability distributions for the duration time of the epidemic, the number of susceptible people, resistent people and dead people after
the epidemic has taken place. Cumulative probabilities on the vertical axis are interpreted as probabilities of the event that fewer than a certain fraction of
individuals on the horizontal axis fall within the class of died, resistent, and susceptible people. The last diagram contains the duration time of the epidemic
and its cumulative probability distribution that the epidemic last shorter than the time on the horizontal axis. The results were obtained by 1000 Monte Carlo
samples of the probabilistic model using the input values from Table 1 }
\label{fig:6}       
\end{figure}

The results have been plotted in Figure \ref{fig:4}. It can be seen that the number of infected people remains small and here the number of people that is
infected is smaller than the number of people that recover from the virus. Hence the number of resistent  and dead people remains very small. This result
indicates that lockdown scenarios (social distancing) can help reducing the number of infected people and ill people. This reduction could also correspond to
having people wearing face masks.
\subsection{Uncertainty Quantification}
\begin{figure}[b]
\sidecaption
\includegraphics[scale=.21]{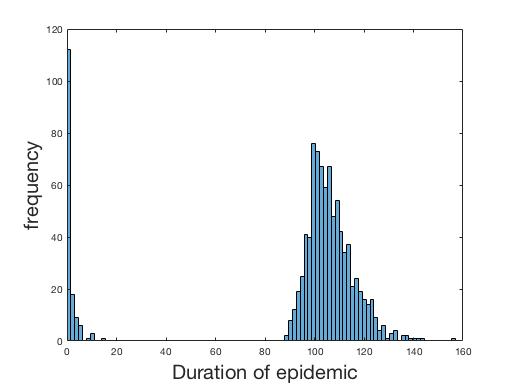}
\includegraphics[scale=.21]{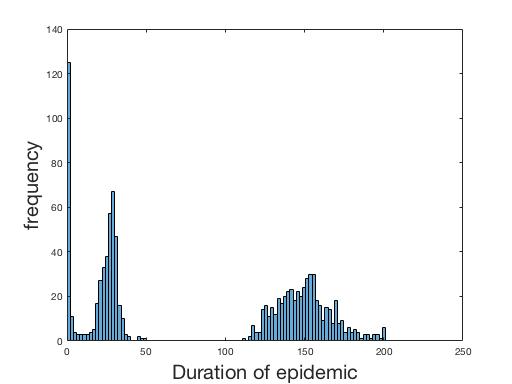}
\includegraphics[scale=.21]{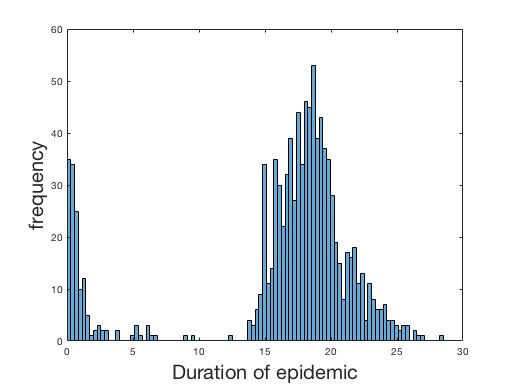}%
%
\caption{Histograms for the duration time of the epidemic for three scenarios: no lockdown, 'weak lockdown' and 'firm lockdown' during a the period between
10 and 30 days, $T_{ld} = (10,30)$, after first breakout of the epidemic. Left: no lockdown, right: $\tilde{\beta} = 0.1$. All simulations have been taken into account except for outliers
outside the interval $(\hat{\mu} - 3 \hat{\sigma},\hat{\mu} + 3 \hat{\sigma})$ }
\label{fig:7}       
\end{figure}

Since the model is based on random principles, we present the results from twenty runs of the model to show the variability in the results
using the basic input values from Table 1. For the sake of illustration, we do not incorporate
lockdown scenarios. The results have been plotted in Figure \ref{fig:5}. Despite the variation in the modelling results, all curves more or less reproduce the 
same trends. The epidemic is over after about a little more than a hundred days. The number of casualties is ranging between 25 and 30 \%, whereas the
portion that becomes resistent is in the order of 65 -- 70\%. The number of people that is not infected at all lies in the order of 5 \%. Once again, we remark 
that we have just been modelling a hypothetic scenario. The calibration of the model to real outbreaks is beyond the scope of the current paper. 
\\[2ex]
Using the Monte Carlo simulations with the model, we show how the model can be used to predict likelihoods of events such as that the length of the time
interval that the epidemic exceeds a certain interval, the total fraction of people that have not been exposed exceeds a certain number, the total number of
deaths exceeds a certain number or that the number of people that become resistent to the virus. In this simulation, 1000 Monte Carlo samples have been
run using the input data set from Table 1. In order to obtain an indication of the Monte Carlo error, we estimate the relative Monte Carlo error of the stochastic 
variable $x$ by
$$
\hat{MCE}_N = \frac{s^f_N}{\hat{x}^N\sqrt{N}} = \sqrt{\frac{\displaystyle{\sum_{j=1}^N (x^{(j)} - \hat{x}^N)^2}}{(\hat{x}^N)^2 ~N(N-1)}}.
$$
Here $N$, $s_N$, $x^{(j)}$, $\hat{x}^N$, respectively, represent the number of samples, sample standard deviation, value of $x$ at sample $j$, and 
the sample mean of $x$ after taking $N$ samples. In our simulations where we carried out 1000 samples, the relative Monte Carlo Error was about
0.02. In Figure \ref{fig:6}, we show Monte Carlo estimates for the cumulative probability distribution for
the fractions of died, resistent and susceptible people. It can be seen that the most simulations indicate a fraction of about 0.275, 0.685, and 0.05 with 
some variation for the dead, resistent, and susceptible people. Furthermore, the duration times vary around 105 days. The cumulative probability densities
that are presented in Figure \ref{fig:6} show the results once the epidemic has lasted for at least two weeks. Since the initial condition was taken as one
infected node only, it happened in a few cases that the one node transferred from infected to resistent before contamination to other individuals could happen.
A probabilistic model is able to predict such rare cases. These cases, and further cases in which the epidemic lasted shorter than two weeks have been ignored
in Figure 6. For completeness, these cases were incorporated in Figure \ref{fig:7}, where histograms for the duration time are presented for several (lockdown) 
scenarios. 

\begin{figure}[b]
\sidecaption
\includegraphics[scale=.33]{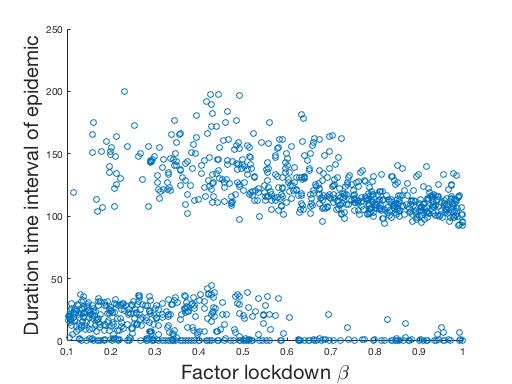}
\includegraphics[scale=.33]{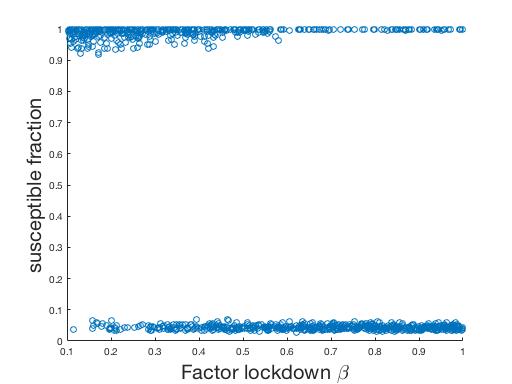} \\
\includegraphics[scale=.33]{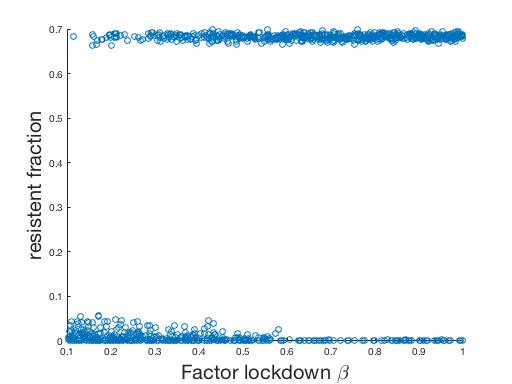}
\includegraphics[scale=.33]{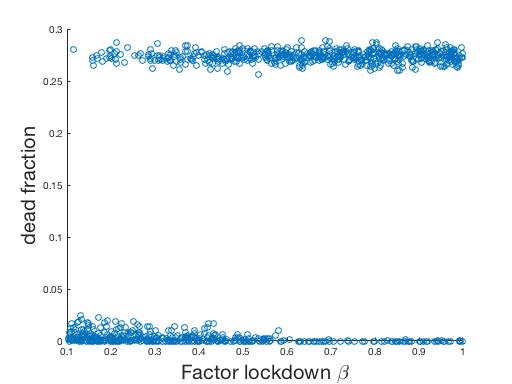}
%
%
\caption{Scatter plots from Monte Carlo simulations for various output parameters as a function of the intensity of the lockdown, $\hat{\beta}$.
Scatter plots are shown for the duration of the epidemic, and the fractions of susceptible, resistent and dead people 
at the moment that the epidemic is over The results were obtained by 1000 Monte Carlo
samples of the probabilistic model }
\label{fig:8}       
\end{figure}

Figure \ref{fig:7} shows histograms for various lockdown scenarios where the intensity of the lockdown policy has been varied. It can be seen that the number of
occurrences that the epidemic is over within 10 days remains the same (note the scale of the horizontal axis), which is according to expectations since the 
lockdown strategy is applied after 10 days after viral outbreak. Further, it can be seen that a fierce reduction of interpersonal contact, which decreases the 
likelihood of contamination, 
significantly shortens the duration of the epidemic.  
All scenarios predicted that the epidemic was over within this lockdown period, except for one outlier,
which has not be shown here. A fierce reduction of inter-personal contact reduces the duration of the epidemic, and reduces the number of casualties and
resistent (not shown here) dramatically. A weak lockdown, however, increases the duration of the epidemic.
\begin{figure}[b]
\sidecaption
\includegraphics[scale=.21]{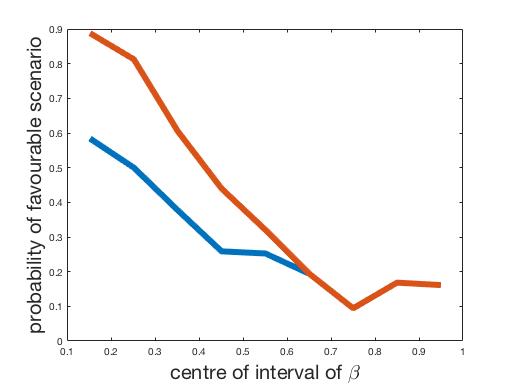}
\includegraphics[scale=.21]{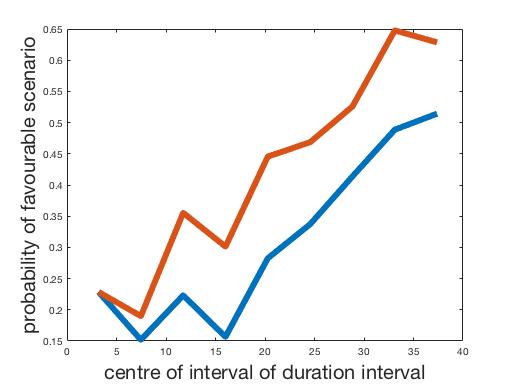}
\includegraphics[scale=.21]{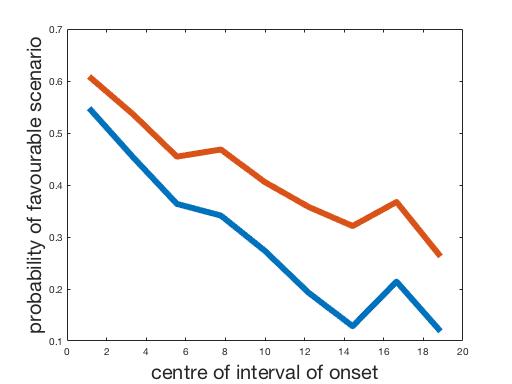}
%
%
\caption{Probability that a given severeness of the lockdown strategy leads to a positive scenario. The horizontal axis gives $\hat{\beta}$, whereas
the vertical axis gives the likelihood of a positive scenario in terms of duration of the epidemic and number of casualties}
\label{fig:9}       
\end{figure}
\\[2ex]
Subsequently a batch of 1000 Monte Carlo simulations were run in which the input parameters have been sampled from statistical distributions:
\begin{equation}
\tilde{\beta} \sim U(0.1,1), ~ \min \{t \in T_{ld}\} \sim U(0,20), ~ \max \{t \in T_{ld}\} \sim U(20,40),
\end{equation}
The results are shown in Figure \ref{fig:8}. The duration of the epidemic is shown versus the severity of lockdown, as well as the fraction of people that 
are susceptible (who have not been infected) and the resistent (who recovered, but may have permanent health issues), and the dead versus the
severity of the lockdown.
 All plots show a bifurcation, which results from the fact that in some runs the epidemic was over before it could
spread significantly. This may happen since both the recovery rate and spreading rate are modelled on the basis of stochastic processes. In reality, this may
also happen if the first receiver of the virus recovers before (s)he spreads the virus to other people. Regarding the bifurcating behaviour, in which there is a 
'favourable' and 'non-favourable' branch (short duration versus long duration, large susceptible end fraction versus small susceptible end fraction,
small resistent/dead end fraction versus large resistent/dead fraction). It can be seen that a more severe lockdown (smaller value of $\hat{\beta}$ results into
a larger number of occurrences in the favourable branch, which suggests a larger likelihood that a favourable scenario occurs in terms of smaller duration 
interval, smaller amount of dead and recovered people, and a larger number of people that did not develop any symptoms. It can also be seen that a very strict 
lockdown (reduction to 10 \% of the number contacts between people) no occurrences in the non-favourable branch were observed, which means that this will make
epidemic die out rapidly with only few casualties.

This has been worked out in Figure \ref{fig:9}, where the likelihood of a positive scenario in terms of number of casualties and duration of epidemic has been 
evaluated for different values of the severeness of the lockdown ($\hat{\beta}$). The probabilities have been estimated using the Maximum Likelihood
Principle. It is clear to see that a more severe lockdown (large reduction of inter-personal contacts) increases the likelihood that the epidemic will last during
a relatively short time-interval. The same can be observed for the number of deaths.

\begin{figure}[b]
\sidecaption
\includegraphics[scale=.33]{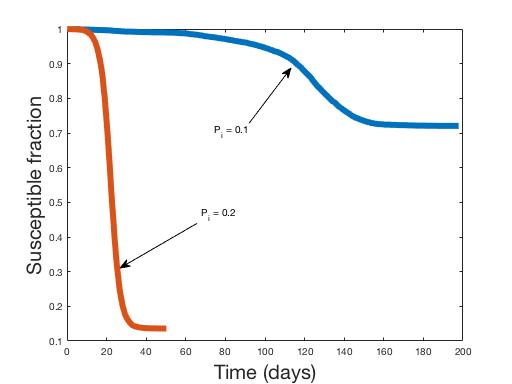}
\includegraphics[scale=.33]{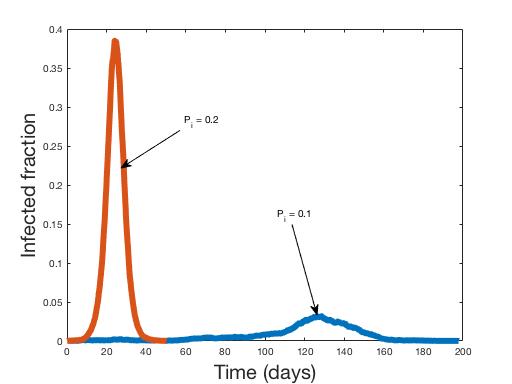} \\
\includegraphics[scale=.33]{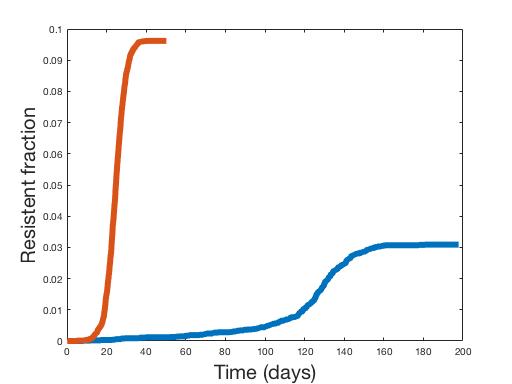}
\includegraphics[scale=.33]{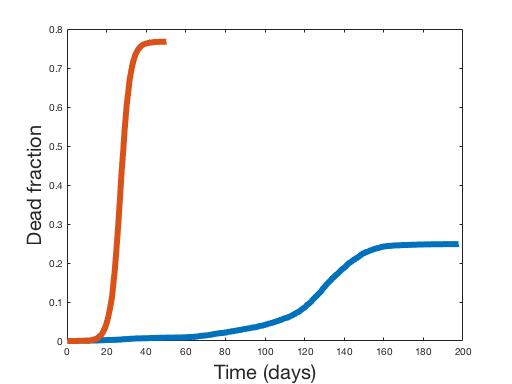}
%
%
\caption{Evolution curve of the epidemic in terms of the fraction of susceptible, infected, resistent and dead people as a function of time for a 
random network model. Curves are shown for a inter-personal connection likelihood of $p_I = 0.1$ and $p_I = 0.2$ }
\label{fig:10}       
\end{figure}

The final simulations that are presented here involve a variation of the number of inter-personal contacts. Here we incorporate a likelihood $p_I$ to each inter-nodal
connection. The random graph that we obtain is obtained by in principle connecting all nodes to one another. Subsequently, a random process is applied
by sampling $\xi$ from a standard uniform distribution, that is $\xi \sim U(0,1)$. Subsequently, we set for the initial adjacency matrix between node $i$ and $j$:
\begin{equation}
a_{ij} = 
\begin{cases}
1, \text{ if } \xi < p_I, \\
0, \text{ if } \xi \ge p_I.
\end{cases}
\end{equation}
This matrix determines the set of connected indices to any point $i$ $\mathcal{N}_j$.
This matrix is again updated by the random fluctuations in inter-personal contacts, by acting
$$
a(i,j) \sim U(0,1), \qquad \text{ if } j \in \mathcal{N}_i.
$$
This option allows a flexible number of contacts between individuals and allows to have isolated subcommunities, in particular of $P_I$ is small. We show some
preliminary results for $n = 10000$ with $p_I = 0.1$ and $p_I = 0.2$. It can be seen that the duration of the epidemic shortens significantly upon having a larger 
probability for connections, however, the number of casualties and recovered people increase as well for a larger probability. The number of susceptible people
decreases of course with an increased likelihood for inter-personal connections. From this, it can be seen easily that a closed society is beneficial for
preventing epidemic casualties. Some further computations were done for a lower likelihoods of 0.05 and lower, then the current parameter set gave no 
further proliferation of the virus. The results are shown in Figure \ref{fig:10}.
%
\section{Conclusions}
We implemented a spatial Markov Chain model using different network topologies for the progression of an epidemic using stochastic principles. 
We deviced the model such that an uncertainty assessment, in the sense that the likelihood of different scenarios is computed, can be carried
out. The model incorporates various lockdown scenarios and can be used to predict the time-evolution of epidemics under various lockdown 
strategies.


\begin{thebibliography}{9}%
%
%

\bibitem{politico} https://www.politico.eu/article/europes-coronavirus-lockdown-measures-compared/

\bibitem{sir} Kermack, W. O.; McKendrick, A. G.: A Contribution to the Mathematical Theory of Epidemics. Proceedings of the Royal Society A: Mathematical, Physical and Engineering Sciences. 115 (772): 700--721 (1927)

\bibitem{Getz} Getz, W.M., Salter, R., Muellerklein, O., Yoon, H.S., Tallam, K.: Modelling epidemics: a primer and numerus model builder implementation.
Epidemics. 25: 9 --19 (2018)

\bibitem{Allen} Allen, L.J.S.: A primer on stochastic epidemic models: formulation, implementation and numerical analysis.
Infectious Disease Modelling. 2: 128--142 (2017)

\bibitem{Trapman} Pellis, L., Ball, F., Bansal, S., Eames, K., House, T., Isham, V., Trapman, P.: Eight challenges for network epidemic models.
Epidemics. 10: 58--62 (2015)

\bibitem{Polonen}
Vermolen, F.J., P\"ol\"onen, I.: Uncertainty quantification on a spatial Markov-chain model for the progression of skin cancer.
J. Math. Biol. 80, 545--573 (2019)


\end{thebibliography}
\end{document}